\documentclass[%
reprint,
preprintnumbers,
nofootinbib,
amsmath,amssymb,
aps,
prl,
]{revtex4-1}

\usepackage{graphicx}
\usepackage{bm}
\usepackage{bbold}
\usepackage{multirow}
\usepackage{amsmath,amssymb,mathtools}
\usepackage{epsfig}
\usepackage[usenames,dvipsnames]{color}
\usepackage{xcolor}
\usepackage{xspace}
\usepackage{slashed}
\definecolor{darkblue}{rgb}{0,0,0.5}
\definecolor{darkgreen}{rgb}{0.0,0.5,0.2}
\definecolor{darkred}{rgb}{0.6,0,0}
\usepackage[pdfstartview=XYZ,
bookmarks=true,
colorlinks=true,
linkcolor=darkred,
urlcolor=blue,
citecolor=darkgreen,
pdftex,
bookmarks=true,
breaklinks=true,
linktocpage=true, 
hyperindex=true
]{hyperref}
\usepackage[capitalise]{cleveref}
\usepackage{orcidlink}

\newcommand{\Ux}{$U(1)_{X}$\xspace}

\newcommand{\C}[1]{\mathcal{#1}}

\begin{document}

\preprint{IPPP/22/40}

\title{A Consistent Theory of Kinetic Mixing and the Higgs Low-Energy Theorem}

\author{Martin Bauer\,\orcidlink{0000-0003-4079-6368}}
\email{martin.m.bauer@durham.ac.uk}
\affiliation{Institute for Particle Physics Phenomenology, Durham University, Durham DH1 3LE, United Kingdom} 
\author{Patrick Foldenauer\,\orcidlink{0000-0003-4334-4228}}
\email{patrick.foldenauer@durham.ac.uk}
\affiliation{Institute for Particle Physics Phenomenology, Durham University, Durham DH1 3LE, United Kingdom}

\date{\today}

\begin{abstract}
Extensions of the Standard Model of particle physics with new Abelian gauge groups allow for kinetic mixing between the new gauge bosons and the hypercharge gauge boson, resulting in mixing with the photon. In many models the mixing with the hypercharge gauge boson captures only part of the kinetic mixing term with the photon, since the new gauge bosons can also mix with the neutral component of the $SU(2)_L$ gauge bosons. We take these contributions into account and present a consistent description of kinetic mixing for general Abelian gauge groups both in the electroweak symmetric and the broken phase.
We identify an effective operator that captures the kinetic mixing with $SU(2)_L$
and demonstrate how renormalisable contributions arise if the charged fields only obtain their masses from electroweak symmetry breaking. For the first time, a low-energy theorem for the couplings of novel Abelian gauge bosons with the Standard Model Higgs boson is derived from the one-loop kinetic mixing amplitudes. 
\end{abstract}

\maketitle

\emph{\textbf{Introduction.}}~\label{sec:intro}\,Extensions of the Standard Model (SM) with an additional Abelian gauge group allow for a unique interaction with SM particles by kinetic mixing with the photon.  
The operator
\begin{align}\label{eq:kinmix1}
\mathcal{L}\supset\ -\frac{\epsilon_A}{2} F_{\mu\nu} X^{\mu\nu}  \,,  
\end{align}
mediates the mixing between the field strength tensors $F^{\mu\nu}$ and $X^{\mu\nu}$ of the electromagnetic gauge group $U(1)_\text{em}$ and the additional gauge group $U(1)_X$, respectively~\cite{Okun:1982xi,Holdom:1985ag}. For two massive gauge bosons there is an additional operator that can mediate mass mixing, \emph{e.g.} for the $U(1)_X$ gauge boson $X$ and the $Z$ boson,
\begin{align}
\mathcal{L}\supset{m_{ZX}^2}\, Z_\mu X^\mu \,.  
\end{align}
The presence of these operators in the broken phase of the electroweak theory have implications for the $U(1)_X$ charges of SM fields in the unbroken phase. For mass mixing to be present the SM Higgs field needs to be charged under $U(1)_X$ or an additional scalar needs to carry charges of the SM gauge group and $U(1)_X$. Kinetic mixing requires either a tree-level mixing between the hypercharge and the \Ux gauge boson, or the presence of fields charged under both $U(1)_Y$ and $U(1)_X$ such that the operator
\begin{align}\label{eq:hyper_mix}
    \C{L} \supset - \frac{\epsilon_{B}}{2} \, B_{\mu\nu} X^{\mu\nu} \,, 
\end{align}
is generated at loop level~\cite{Degrassi:1989mu, delAguila:1995rb}. In this paper we focus on one important exception to the latter argument. Mixing between the photon and the $X$ boson can also be generated if fields carry $U(1)_X$ charges and are $SU(2)_L$ multiplets. In this situation no renormalisable kinetic mixing operator is present in the symmetric phase, but loop effects induce kinetic mixing between the neutral component $W^3$ of $SU(2)_L$ and the $X$ boson that generate the operator \eqref{eq:kinmix1}.
In particular, extensions of the SM in which SM fermion doublets carry $U(1)_X$ charges inevitably give rise to this contribution. This class of models is of great phenomenological interest and includes extensions of the SM with a gauged baryon lepton number difference $U(1)_{B-L}$, gauged lepton family number differences   $U(1)_{L_{\mu}-L_e}$,  $U(1)_{L_{e}-L_\tau}$ and $U(1)_{L_{\mu}-L_\tau}$~\cite{Fayet:1990wx, Foot:1990mn,He:1990pn, He:1991qd,Leike:1998wr,Ma:2001md,Baek:2001kca,Langacker:2008yv, Heeck:2011md, Heeck:2011wj,Araki:2012ip,Heeck:2014zfa,Ilten:2018crw,Arcadi:2018tly, Bauer:2018onh,Foldenauer:2018zrz,Asai:2018ocx,Escudero:2019gzq,Bauer:2020itv}, as well as combinations thereof~\cite{Chang:2000xy,Bian:2017rpg,Bian:2017xzg, Kamada:2018kmi}. In this Letter 
we calculate the contributions from mixing between the $X$ and the neutral $SU(2)_L$ gauge boson in a general form, which apart from few exceptions~\cite{Barello:2015bhq,Barducci:2021egn,Rizzo:2022jti} have been neglected or omitted in the literature,
and derive consistent expressions at the tree and loop level.
We further use these expressions to provide the first formulation of low-energy theorems for Higgs couplings with the $X$ boson that are relevant for Higgs decays in any theory involving kinetic mixing. We point out important differences with the low-energy theorems for the SM mixing between the photon and the $Z$ boson~\cite{Kniehl:1990mq,Kniehl:1995tn,Carena:2012xa}.\\[.1cm]

\emph{\textbf{Tree-level mixing.}}~\label{sec:tree_mix}\,In the special class of purely secluded hidden photon models, in which the SM gauge group is extended by an Abelian group \Ux under which no fields are charged, it is sufficient to consider the mixing term \eqref{eq:hyper_mix} between the new \Ux boson and the SM hypercharge field. 

However, if there are either new fields charged under both \Ux and the SM $SU(2)_L$ or any of the SM fermion doublets are charged under \Ux, there will be an effective operator
\begin{align}\label{eq:o_wx}
    \mathcal{O}_{WX} = \frac{c_{WX}}{\Lambda^2}\, H^\dagger \sigma^i H \, W^i_{\mu\nu} X^{\mu\nu} \,,
\end{align}
in which $H$ denotes the Higgs doublet and $W_{\mu\nu}^i$ the $SU(2)_L$ field strength tensor.~\footnote{Note that a similar operator is present in models in which the SM is extended by an $SU(2)_L$ triplet scalar~\cite{Brummer:2009oul,Arguelles:2016ney,Rizzo:2022jti}.}

The scale $\Lambda$ will be identified with the mass of the heavy states and the operator vanishes in the limit ${\Lambda\to \infty}$. If these states obtain their masses solely through the Higgs mechanism the scale $\Lambda$ is identified with the vacuum expectation value (vev) of the Higgs field, resulting in a dimensionless coefficient of the $W^3-X$ mixing operator $\mathcal{O}_{WX}$~\cite{Pierce:2006dh}. 
We can make this explicit by replacing the Higgs doublet in \eqref{eq:o_wx} by $H=1/\sqrt{2}\,(0, v+h)^T$, so that mixing between the neutral  $SU(2)_L$ gauge boson and the hidden photon, 
\begin{align}
    \mathcal{L}
    \supset &  - \frac{\epsilon_W}{2} \,  W^3_{\mu\nu} X^{\mu\nu}\label{eq:o_wx_mix}\,,
\end{align}
is generated with a coefficient $\epsilon_W= c_{WX}\, v^2/\Lambda^2$.

Including both \eqref{eq:hyper_mix} and \eqref{eq:o_wx_mix}, we can write the relevant part of the SM and $U(1)_X$ Lagrangian as
\begin{align}\label{eq:w3_treemix}
\mathcal{L} =&- \frac{1}{4}
(B_{\mu\nu}, W^3_{\mu\nu}, X_{\mu\nu})
\left(\begin{matrix}
    1 & 0 & \epsilon_B\\
    0 & 1 & \epsilon_W  \\
    \epsilon_B & \epsilon_W  & 1
    \end{matrix}\right)
    \left(\begin{matrix}
    B^{\mu\nu}\\
    W^{3\mu\nu}\\
    X^{\mu\nu}
    \end{matrix}\right)\notag \\
    &-g' j^Y_\nu B^{\nu}- g\, j^3_\nu W^{3\nu}-g_x j^x_\nu X^{\nu}\,,
\end{align}
where $g' j^Y_\nu,\, g\, j^W_\nu $ and $g_x j^x_\nu$ denote the gauge couplings and currents of the hypercharge $U(1)_Y$, the weak $SU(2)_L$, and $U(1)_X$, respectively. It is straightforward to write the Lagrangian in the broken electroweak phase~\cite{Bauer:2018onh}
\begin{align}
\mathcal{L}_\text{em+Z}
=&- \frac{1}{4}
(F_{\mu\nu}, Z_{\mu\nu}, X_{\mu\nu})
\left(\begin{matrix}
    1 & 0 & \epsilon_A \\
    0 & 1 & \epsilon_Z \\
    \epsilon_A & \epsilon_Z & 1
    \end{matrix}\right)
    \left(\begin{matrix}
    F^{\mu\nu}\\
    Z^{\mu\nu}\\
    X^{\mu\nu}
    \end{matrix}\right)\notag\\
    &  - e\, j_\nu^\text{em} A^{\nu} -\frac{g}{c_w} j_\nu^ZZ^{\nu}-g_x j_\nu^xX^{\nu}\,,
\end{align}
where $Z^{\mu\nu}$ is the field strength tensor of the $Z$ boson, $s_w, c_w$ are the sine and cosine of the Weinberg angle $\theta_W$ and the currents and mixing parameters are related by
\begin{align}
j_\nu^\text{em}= \,&g'\,c_w\, j^Y_\nu + g\,s_w\, j^3_\nu\,,\quad \epsilon_A=c_w\, \epsilon_B + s_w\, \epsilon_W\,,\\
j_\nu^Z=-&g'\,s_w\, j^Y_\nu + g\,c_w\, j^3_\nu\,,\quad  \epsilon_Z=-s_w\, \epsilon_B + c_w\, \epsilon_W\,.
\end{align}
If we want to enforce the absence of any $SU(2)_L$-breaking mixing term between the $W^3$ and the $X$ boson, we have to require that $\epsilon_W=0$ (at tree level) and~\cite{Pospelov:2008zw,Bjorken:2009mm,Fabbrichesi:2020wbt}
\begin{equation}\label{eq:simple_mix}
    \epsilon_Z = - \frac{s_w}{c_w}\, \epsilon_A\,.
\end{equation}
In this Letter we focus on \Ux extensions under which (some of) the SM fermions are charged while the SM Higgs remains uncharged. 
In the next section, we will show explicitly how the operator $\mathcal{O}_{WX}$ is generated at the one-loop level in these models.\\[.3cm]

\emph{\textbf{Loop mixing.}}~\label{sec:model}\,In the absence of tree-level mixing, the coefficients $\epsilon_B$ and $\epsilon_W$ can be generated at loop level. We consider the contribution of fermions $f$ charged under $U(1)_X$, $U(1)_Y$ and $SU(2)_L$.
Although $U(1)$ gauge bosons couple to conserved vector currents, for the sake of generality, we give the result for vector $v^f_{i}=Q^{f}_{Li}+Q^f_{Ri}$ and axial-vector couplings $a^f_{i}=Q^{f}_{Li}-Q^f_{Ri}$ for both hypercharge $i=B$ and the hidden photon $i=X$ in terms of the charges of the left- and right-handed fermions, $Q^f_L$ and $Q^f_R$.~\footnote{Note that these couplings are defined in the mass eigenbasis and  differ from the expressions in the interaction eigenbasis for fermions with both vector- and Yukawa mass terms.} The relevant vacuum polarisation amplitudes are given in $d=4-2\varepsilon$ dimensions by 
\begin{align}\label{eq:pi2general}
      \Pi^{\mu\nu}_{BX} \!&=\! \frac{ g' g_x}{8\pi^2}\sum_f \int^1_0\!\! dx 
\bigg[ a^f_{B}\,a^f_{X}\, m^2_f\, g^{\mu\nu} \\
&-(v^f_{B}\,v^f_{X} + a^f_{B}\,a^f_{X})\,x(1-x) \, [  g^{\mu\nu} q^2 - q^\mu q^\nu ] \bigg] C_\epsilon\notag\,, \\
\label{eq:mixw3}
\Pi^{\mu\nu}_{W^3X} \!&=\! \frac{ g\,g_x}{8\pi^2}\sum_f \int^1_0\!\! dx  \,T^f_3
\bigg[ a^f_{X}\, {m^2_f}\, g^{\mu\nu}\\
&-\big(v^f_{X} + a^f_{X}\big)\,x(1-x) \, [ g^{\mu\nu} q^2 - q^\mu q^\nu  ] \bigg] C_\epsilon\notag \,,
\end{align}
where 
\begin{equation}
C_\epsilon=    \frac{2}{\varepsilon}- \gamma_E + \log\left(\frac{\mu^2}{m_f^2-x(1-x)q^2}\right) +\mathcal{O}(\epsilon)\,.
\end{equation}
The sum in the expression~\eqref{eq:mixw3} runs over the degrees of freedom of complete $SU(2)_L$ multiplets. First we focus on the kinetic mixing terms in the second lines of~\eqref{eq:pi2general} and~\eqref{eq:mixw3}. Since all degrees of freedom of each multiplet carry the same \Ux charges, the divergence in \eqref{eq:mixw3} automatically cancels within each multiplet 
\begin{equation}
  \sum_f (v_X^f+a_X^f)\, T^f_3 =0 \,.
\end{equation}
For kinetic mixing between two $U(1)$ gauge bosons the divergence in \eqref{eq:pi2general} only cancels for charges with
\begin{equation}\label{eq:b_cancel}
   \sum_f (v_B^f v_X^f+a_B^fa_X^f) =0 \,.
\end{equation}
In this special case the tree-level counterterm is not needed and the $U(1)_X$ can be embedded in a direct product ${SU(3)_C\times SU(2)_L\times U(1)_Y \times G_X}$ with a Non-Abelian gauge group $G_X$ and a spontaneous symmetry breaking  $G_X \to U(1)_X$~\cite{Bauer:2018onh}. This is the case for the gauged lepton family number differences    $L_\mu-L_e $, $L_e-L_\tau$ or $L_\mu-L_\tau$, but not for $B-L$.  
Note that in the case $a_x^f=0$ and denoting the SM hypercharge for the fermion multiplet $f$ by $Y_f$, the condition~\eqref{eq:b_cancel} can be written as $\sum_f\, Y_f\, v_x^f = 0$, as discussed in~\cite{delAguila:1995rb,Babu:1996vt, Loinaz:1999qh,Rizzo:2018vlb}. 

Contributions to mass mixing, like the ones in the first lines of~\eqref{eq:pi2general} and~\eqref{eq:mixw3}, only arise if the fermions are chiral with respect to $U(1)_X$ and the hidden photon has axial-vector couplings. In order to construct renormalisable Yukawa couplings in any model with two $U(1)$ gauge groups with different charges for left- and right-handed fermions one needs to introduce a symmetry-breaking scalar field charged under both gauge groups. This scalar gives rise to a tree-level mass mixing between the two gauge bosons, providing a counterterm for the divergence in \eqref{eq:pi2general}. The contribution to mass mixing with the neutral $SU(2)_L$ gauge boson is always finite.

Matching the amplitudes~\eqref{eq:pi2general} \& \eqref{eq:mixw3} to the kinetic mixing operators in~\eqref{eq:hyper_mix} and~\eqref{eq:o_wx_mix}, and defining the integral
\begin{align}
L_f=    \int_0^1 \!\!dx \, x(1-x)\, \log\left(\frac{\mu^2}{m_f^2-x(1-x)q^2}\right)\,,
\end{align}
we find for the kinetic mixing coefficients (taking ${v^f_X = 2\,Q^f_X}$ and $a^f_X = 0$)
\begin{align}
    \epsilon_{B} &= \frac{ g' g_x}{8\pi^2} \sum_f \big[ Y^f_L\! +\!Y^f_R\big] Q^f_X \,L_f\,, \\
    \epsilon_{W} &= \frac{ g\, g_x}{4\pi^2}\, \sum_f T^f_3\, Q^f_X \,L_f\,.
\end{align}
The $\mu^2$-dependence cancels between the contributions from all charged fermions. In the broken basis, the coefficients for the mixing between the hidden photon and the photon or $Z$ boson read
\begin{align}
    \epsilon_A &= 
     \frac{e\, g_x}{8\pi^2} \!\sum_f  \left[ 2T^f_3+ Y^f_L +Y^f_R\right] Q^f_X L_f\,,\\[2pt]
    \epsilon_Z  
    &= \frac{g_Z g_x}{8\pi^2}\!\sum_f \Big[2T^f_3 -s_w^2 \left(2T^f_3+Y^f_L +Y^f_R\right) \Big] Q^f_X L_f \,.
\end{align}

We illustrate the importance of the operator $\mathcal{O}_{WX}$ for the correct matching with the example of a vectorlike $SU(2)_L$ fermion doublet  
$N=(N^u, N^d)$ with hypercharge $Y^N=0$ and two vectorlike singlets $\chi$ and $\eta$ with hypercharge $Y^\chi=+1$ and $Y^\eta=-1$. All fields carry $U(1)_X$ charges $Q_X=1$ and have mass terms
\begin{align}\label{eq:lag_vec}
    \mathcal{L}\supset M\, (\bar N N +\bar \chi \chi+ \bar \eta \eta) +y_u\, \bar N \tilde H \chi+y_d\, \bar N  H \eta\,,
\end{align}
where the choice of equal vectorlike fermion masses for the doublet and singlets simplifies the calculation but does not change the conclusions. Upon electroweak symmetry breaking the mass eigenstates can be written as $N^{1,2}=1/\sqrt{2}(N^u \pm \chi)$ and $N^{3,4}=1/\sqrt{2}(N^d \pm \chi)$ with masses $m_{1,2}=M\pm y_uv$ and $m_{3,4}=M\pm y_dv$, respectively. There are now four diagrams contributing to $\epsilon_W$ with $N^1$, $N^2$, $N^3$ and $N^4$ running in the loop. Adding up the different contributions for $q^2\to 0$ one obtains the finite expression
\begin{align}
\epsilon_W&= \frac{gg_x}{192\pi^2} \log\left(\frac{m_3^2m_4^2}{m_1^2m_2^2}\right)\,.
\end{align}
In the limit $v\ll M$, this expression reduces to
\begin{align}\label{eq:expansion}
    \epsilon_W&= \frac{gg_x}{96\pi^2}\left[(y_u^2-y_d^2)\frac{v^2}{M^2}+\mathcal{O}\bigg(\frac{v^4}{M^4}\bigg)\right]\,, 
\end{align}
as expected from the operator \eqref{eq:o_wx} with $\Lambda=M$. There is no logarithm in the expansion \eqref{eq:expansion}, because the operator \eqref{eq:o_wx_mix} is a measure of $SU(2)_L$ breaking and for multiplets with degenerate masses the mixing term ${\epsilon_W\propto \text{Tr}[T_3]=0}$ vanishes independent of the representation. 
In the opposite limit $v \gg M$, only a logarithm is present, 
\begin{align}
\epsilon_W&= \frac{gg_x}{96\pi^2} \log\left(\frac{y_d}{y_u}\right)\,.
\end{align}
The dimension-6 operator $O_{WX}$ captures the $SU(2)_L$ contributions to kinetic mixing of the $X$ boson with $W^3$, and together with the contribution from $X$ mixing with the hypercharge gauge boson this reproduces the mixing with the photon (and the $Z$ boson) in the effective theory with only $U(1)_\text{em}$ (and using the $Z$ boson couplings).
Examples of theories in which the contribution to \eqref{eq:o_wx_mix} arises at the renormalisable level are gauge groups that are anomaly-free with charged SM fields alone. For example, in the case of $U(1)_{L_\mu-L_\tau}$ one has in the limit $q^2\to 0$
\begin{align}
\epsilon_B &= \frac{g' g_{\mu-\tau}}{24\pi^2}\left[3\log\left(\frac{m_\mu}{m_\tau}\right) +\log\left(\frac{m_{\nu_\mu}}{m_{\nu_\tau}}\right)\right]\,,\\
 \epsilon_W &= \frac{g g_{\mu-\tau}}{24\pi^2}\left[\log\left(\frac{m_\mu}{m_{\nu_\mu}}\right) -\log\left(\frac{m_{\tau}}{m_{\nu_\tau}}\right)\right]\,.
\end{align}

The fact that loops of fermions that obtain their mass from the electroweak scale can generate contributions to operators that are effectively of dimension $d-2$, while the same diagrams generate contributions to dimension $d$ operators if the fermions have vector masses $M \gg v$ is not unique to kinetic mixing. It has been discussed for the case of di-higgs production $gg\to hh$~\cite{Pierce:2006dh}, the decay of heavy pseudoscalar resonances into Higgs and $Z$ bosons~\cite{Bauer:2016ydr,Bauer:2016zfj} and for exotic Higgs decays into $Z$ bosons and axions~\cite{Bauer:2017nlg, Bauer:2017ris}.\\[.2cm]

\begin{figure*}
    \centering
    \includegraphics[width=.95\textwidth]{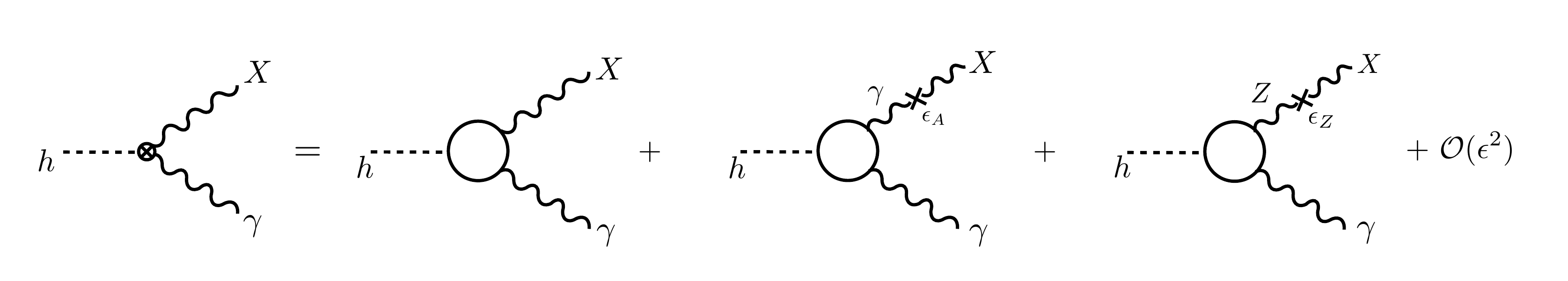}
    \caption{Diagrammatic representation of the different contributions from fermion loops to the amplitude $\mathcal{M}(h \to \gamma X)$ up to quadratic order in the mixing parameters $\epsilon \approx \epsilon_A \approx \epsilon_Z$.}
    \label{fig:diagrams}
\end{figure*}

\emph{\textbf{Low-energy theorems.}}~\label{sec:let}\,Using our general result for the contribution to the vacuum polarisation of two vector bosons due to charged fermions in~\eqref{eq:pi2general}, we can formulate a low-energy theorem for Higgs couplings to SM photons, the $Z$ and the $X$ boson of a new $U(1)_X$.~\footnote{The corresponding contributions from charged scalars are straightforward to derive.} 
We work in a low-energy theory where we keep the $Z$ boson, the photon and the $X$ boson as the relevant degrees of freedom. All loop-contributions are evaluated at zero momentum transfer $q^2=0$,  such that the full vacuum polarisation amplitude can be defined as
\begin{align}
     \Pi^{\mu\nu}_{ij} & = \Pi_{ij}\, [g^{\mu\nu} p_1\cdot p_2  - p_1^\mu p_2^\nu] + \Delta_{ij} \,  g^{\mu\nu} \,,
\end{align}
with all $\Pi_{ij}\equiv \Pi_{ij}(0) $ and $\Delta_{ij}\equiv \Delta_{ij}(0)$, and
\begin{align}\label{eq:delpi}
    \Pi_{ij} &= - \frac{ g_ig_j}{8\pi^2}  \,\frac{1}{6} \sum_fN_c^f(v^f_{i}\,v^f_{j} + a^f_{i}a^f_j) \log\left(\frac{\mu^2}{m_f^2}\right)\,,\\ \label{eq:deldel}
    \Delta_{ij} &= \frac{ g_ig_j}{8\pi^2} \,\sum_f N_c^f a^f_{i}\,a^f_{j}\, m^2_f\ \log\left(\frac{\mu^2}{m_f^2}\right) \,.
\end{align}
Note that there are also contributions to $\Pi_{\gamma\gamma}, \Pi_{\gamma Z}, \Pi_{ZZ}$ and $\Delta_{ZZ}$  from charged SM bosons in the loop which are not captured by \eqref{eq:delpi} and \eqref{eq:deldel}.
For fermions with both vector masses and masses from Yukawa couplings, the axial vector couplings $a_i^f$ and $a_j^f$ are defined in the fermion mass eigenbasis here. The vector couplings $v_i^f$ and $v_j^f$ are the same in the interaction and mass eigenbasis since they commute with the rotation matrices.
The different mixing contributions can then be written as
\begin{align}
    \mathcal{L} = &- \frac{1}{4}\,
    (F_{\mu\nu}, Z_{\mu\nu}, X_{\mu\nu})
   \left[\begin{pmatrix}
    1 & 0 & \epsilon_A \\
    0 & 1 & \epsilon_Z \\
    \epsilon_A & \epsilon_Z & 1
    \end{pmatrix}
    +
    \boldsymbol{\Pi}
    \right]
    \begin{pmatrix}
    F^{\mu\nu}\\
    Z^{\mu\nu}\\
    X^{\mu\nu}
    \end{pmatrix} \notag \\[2pt]
    & +  \frac{1}{2} \, 
    (A_{\mu},Z_\mu, X_{\mu})\,
    \left[  \mathbf{M}+\mathbf{\Delta} \right]
    \begin{pmatrix}
    A^{\mu}\\
    Z^\mu \\
    X^{\mu}
    \end{pmatrix}\,. \label{eq:loop_lag}
\end{align}
Here, $\epsilon_A$ and $\epsilon_Z$ denote the tree-level contributions to kinetic mixing, and the mass and mixing matrices are given by  $\boldsymbol{M}=\text{diag}(0,m_Z^2,m_X^2)$, and
\begin{align}\label{eq:loop_matr}
    \boldsymbol{\Pi}=
    \begin{pmatrix}
    \Pi_{\gamma\gamma} & \Pi_{\gamma Z} & \Pi_{\gamma X} \\
    \Pi_{\gamma Z} & \Pi_{ZZ} & \Pi_{ZX} \\
    \Pi_{\gamma X} & \Pi_{ZX} & \Pi_{XX}
    \end{pmatrix}
    \!,\,\,
    \boldsymbol{\Delta} = \begin{pmatrix}
    0&0 & 0 \\
    0& \Delta_{ZZ} & \Delta_{ZX} \\
    0& \Delta_{ZX} & \Delta_{XX}
    \end{pmatrix}\,.
\end{align}
The tree-level kinetic terms in~\eqref{eq:loop_lag} are diagonalised by the non-unitary transformation
\begin{equation}
  G = 
\begin{pmatrix}
 1 & 0 & -\displaystyle{\frac{\epsilon _A}{ \sqrt{1-\epsilon _A^2-\epsilon _Z^2}}} \\[8pt]
 0 & 1 & -\displaystyle{\frac{\epsilon _Z}{ \sqrt{1-\epsilon _A^2-\epsilon _Z^2}}} \\[8pt]
 0 & 0 & \displaystyle{\frac{1}{ \sqrt{1-\epsilon _A^2-\epsilon _Z^2}}} 
\end{pmatrix}\,.
\end{equation}
The matrices in \eqref{eq:loop_matr} have to be rotated as well, and to linear order in $\epsilon_A$ and $\epsilon_Z$ one finds
\begin{align}\label{eq:vac_pol_rot}
         G^T\,&\boldsymbol{\Pi} \,G=\boldsymbol{\Pi} -
    \begin{pmatrix}
    0 & 0 &   \epsilon_A\,\Pi_{\gamma\gamma}+ \epsilon_Z\,\Pi_{\gamma Z} \\[.25cm]
    \cdot &0  &   \epsilon_A\,\Pi_{\gamma Z}+ \epsilon_Z\,\Pi_{ZZ} \\[.25cm]
    \cdot & \cdot & 2\epsilon_A \Pi_{\gamma X} + 2\epsilon_Z \Pi_{Z X}
    \end{pmatrix}\,,  \\[4pt]
    G^T\,\big[&\boldsymbol{M}+\boldsymbol{\Delta}\big] \,G  =\big[\boldsymbol{M}+\boldsymbol{\Delta}\big]-
    \begin{pmatrix}
    0 & 0 &   0\\
    \cdot & 0 &    \epsilon_Z\,(m_Z^2+\Delta_{ZZ}) \\[.25cm]
    \cdot & \cdot &  2\epsilon_Z\, \Delta_{Z X} 
    \end{pmatrix}\,,
\end{align}
where the dots represent the entries obtained from mirroring the matrices at their main diagonal.

Our goal is to derive the amplitudes for the Higgs decays $h\to V_i\,V_j$ with $V_{i,j}=\gamma,Z,X$ from the vacuum polarisation amplitudes via the low-energy theorem~\cite{Ellis:1975ap,Shifman:1979eb,Bergstrom:1985hp,Kniehl:1995tn}
\begin{equation}
    \lim_{p_h\to0} \mathcal{M}(h\to V_i V_j) \to  \frac{\partial}{\partial v} \mathcal{M}(V_i\to V_j).
\end{equation}
Factoring out the gauge boson polarisation vectors from the decay amplitude, 
\begin{align}
\mathcal{M}_{h\to V_i\,V_j} = \mathcal{M}_{h\to V_i\,V_j}^{\mu\nu}\ \epsilon^*_{\mu,\lambda}(p_1)\,\epsilon^*_{\nu,\lambda'}(p_2),
\end{align}
the general Higgs decay amplitudes at low energy are
\begin{align}
     \mathcal{M}_{h\to V_i\,V_j}^{\mu\nu} =&\, {\partial_v} [  G^T\,\boldsymbol{\Pi} \,G]_{ij}\ [p_2^\mu\, p_1^\nu -  p_1\cdot p_2\, g^{\mu\nu}] \notag\\
     &+  {\partial_v} \big[G^T\,\big[\boldsymbol{M}+\boldsymbol{\Delta}\big] \,G \big]_{ij} \  g^{\mu\nu} \,.
    \label{eq:low_en_amp}
\end{align}
For $n=1,\dotsc,N$ multiplets of fermions, which carry the same charge under both gauge groups, one has for the $n^{th}$ multiplet~\cite{Carena:2012xa}
\begin{align}\label{eq:delpi_gen}
    {\partial_v}\, \Pi_{ij}^n=   &\frac{ g_ig_j}{48\pi^2}  N_c^n (v^n_{i}\,v^n_{j} + a^n_{i}a^n_j)\ \partial_v  \log \left( \frac{\mathrm{det} \mathcal{M}^\dagger_n \mathcal{M}_n}{\mu^2} \right)\,,\\
       {\partial_v}\, 
       \Delta_{ij}^n=   &-\frac{ g_ig_j}{8\pi^2}  N_c^n a^{n}_{i}\,a^n_{j}\ \partial_v  \mathrm{Tr} \left[\mathcal{M}^\dagger_n \mathcal{M}_n\log\left( \frac{ \mathcal{M}^\dagger_n \mathcal{M}_n}{\mu^2} \right)\right]\,.
\end{align}
For illustration, we consider again the example of a vectorlike doublet and two singlets charged under $SU(2)_L$ and \Ux given in~\eqref{eq:lag_vec}. In this case the fermion mass matrices read
\begin{equation}
    \mathcal{M}_u =
    \begin{pmatrix}
    M & y_u\, v \\
    y_u\, v & M
    \end{pmatrix} ,\quad   \mathcal{M}_d =
    \begin{pmatrix}
    M & y_d\, v \\
    y_d\, v & M
    \end{pmatrix}, 
\end{equation}
and the one-loop coefficient for the decay $h\to V_i\,V_j$ are 
\begin{align}\label{eq:vec_coeff}
    {\partial_v}\, \Pi_{ij}=&   \frac{ g_ig_j}{12\pi^2} \sum_{k=u,d}  (v^{k}_{i}\,v^k_{j}+a^{k}_{i}\,a^k_{j})\ \frac{ -\,y_k^2\,v}{M^2- (y_k\,v)^2}\,,\\
    {\partial_v}\, \Delta_{ij}=& - \frac{ g_ig_j}{2\pi^2} \sum_{k=u,d}  a^{k}_{i}\,a^k_{j}\ y_k^2\, v \  \Bigg(1+\log\left[\frac{M^2- y_k^2 v^2}{\mu^2}\right] \notag \\
   & \qquad \qquad \qquad \ + \frac{M}{y_k v}\log\left[\frac{M+y_k v}{M- y_k v}\right] \Bigg)\,,
\end{align}
assuming all states are colour singlets, $N^k_c=1$.

In the case of SM fermions in the loop, the expression in~\eqref{eq:delpi_gen} for the Higgs decays into photons, $Z$ and $X$ bosons read
\begin{align} \label{eq:1loop_GX}
    {\partial_v}\, \Pi_{\gamma X}(0) &= \sum_f N_c^f \frac{e\,g_x}{12\,\pi^2 \,v} \,Q_f\,v^f_{X} \,, \\
    {\partial_v}\, \Pi_{Z X}(0) &= \sum_f N_c^f \frac{e\,g_x}{24\,\pi^2 \,v} \,\frac{T_3^f - 2 \,s_w^2\,Q_f}{s_w c_w}\,v^f_{X} \,, \\
    {\partial_v}\, \Pi_{X X}(0) &= \sum_f N_c^f \frac{g_x^2}{24\,\pi^2\, v} \, 
    v^{f2}_{X} \,. \label{eq:1loop_XX}
\end{align}
In these expressions the sum runs over all heavy fermions in the loop with $m_f \gg m_h$  (i.e.~the top quark) and the vector charge is given by  $v^f_\mathrm{em}= 2\, Q_f$ for the SM photon, and by $v^f_{Z}=T_3^f - 2 \,s_w^2\,Q_f$ for the $Z$ boson.
To the best of our knowledge, the low-energy theorem for the decays $h\to \gamma X$, $h\to Z X$ and $h\to XX$ have been derived for the first time in this work.
In particular, it should be noted that the $h\to XX$ decay amplitude is generated purely from the transverse component (i.e.~$\Pi_{XX}$) of the vacuum polarisation since Abelian gauge bosons couple purely vectorially. This is in stark contrast to the low-energy theorem for the SM ${h\to ZZ^*}$ amplitude, which is entirely dominated by the axial coupling of the $Z$ boson and hence is generated from the longitudinal component of the vacuum polarisation~\cite{Kniehl:1990mq,Dawson:1988th}.

As an application of the low-energy theorems, we give the branching fractions for the processes $h\to \gamma X $ and $h\to X X $ in a gauged $U(1)_{B-L}$ model for sub-weak scale mediators ($m_X\ll m_h$). The different diagrams contributing to the amplitude for $h\to \gamma X$ are shown in Fig.~\ref{fig:diagrams}. Under the assumption of the simple tree-level mixing relation~\eqref{eq:simple_mix} these can be expressed by
\begin{align}\label{eq:br_GX}
\mathcal{BR}_{h\to \gamma X}\!&\simeq (0.92\, g_x^2 +6.36g_x\epsilon_A  + 11.01\epsilon^2_A)\,\cdot 10^{-3}\!, \\[3pt]
\mathcal{BR}_{h\to X X}  \!&\simeq g_x^2 (2.5\, g_x^2 - 5.7 \, g_x\epsilon_A + 3.2 \epsilon^2_A)\,\cdot 10^{-3}, \label{eq:br_XX}
\end{align}
to leading order in the gauge coupling $g_x$ and the tree-level kinetic mixing parameter $\epsilon_A$. Similar expressions can be derived for the branching ratios $\mathcal{BR}(h \to ZX)$, $\mathcal{BR}(X\to h\gamma)$, $ \mathcal{BR}(X\to hZ)$, depending on the mass of the $U(1)_X$ boson.
Since in the low-energy limit only the top quark contributes to the fermionic  one-loop expressions in~\eqref{eq:1loop_GX}-\eqref{eq:1loop_XX} as all other SM fermions are much lighter than the Higgs, the branching ratios in~\eqref{eq:br_GX} \& \eqref{eq:br_XX} are universal for all \Ux groups that gauge baryon number $B$. We have taken into account further important contributions to $\Pi_{\gamma\gamma}$ and $\Pi_{\gamma Z}$ from SM $W$ boson loops, which can be found e.g.~in~\cite{Cahn:1978nz,Bergstrom:1985hp,Kniehl:1990mq,Spira:1991tj,Djouadi:1996yq}. We have explicitly checked that including the exact $W$ boson one-loop amplitude amounts to a correction of $\sim12\%$ compared to the low-energy amplitude.
For values of $g_x\sim10^{-4}$ and $\epsilon_A\sim10^{-3}$ we find $\mathcal{BR}_{h\to \gamma X} \sim 10^{-8}$ and $\mathcal{BR}_{h\to X X} \sim 10^{-17}$. The former branching ratio is 4 orders of magnitude smaller than the SM process $h\to \mu\mu$~\cite{LHCHiggsCrossSectionWorkingGroup:2016ypw}, which has been recently observed at the LHC for the first time~\cite{CMS:2020xwi}. Thus, the process $h\to \gamma X$ might be testable at a future collider like the FCC-hh, aiming at collecting up to $\mathcal{O}(10^{10})$ Higgs bosons~\cite{FCC:2018byv}.
\\

\emph{\textbf{Conclusions.}}~\label{sec:conclusions}\,We have derived consistent expressions for kinetic mixing between the gauge boson of a $U(1)_X$ extension of the SM with the neutral component $W^3$ of $SU(2)_L$ that have generally been omitted in the literature. These contributions to kinetic mixing exist only if the SM $SU(2)_L$ is broken and are renormalizable if induced by SM fields in the loop. For new, heavy states charged under \Ux and $SU(2)_L$ these contributions are suppressed by the heavy scale. Therefore, mixing between the $U(1)_X$ and $SU(2)_L$ is most relevant for gauged anomaly-free global symmetries of the SM, such as $U(1)_{B-L}$,  $U(1)_{L_{\mu}-L_e}$, $U(1)_{L_{e}-L_\tau}$, $U(1)_{L_{\mu}-L_\tau}$ and combinations thereof. 
We have further formulated a low-energy theorem expressing the couplings of the Higgs boson to photons, $Z$ and $X$ bosons through the vacuum polarisation amplitudes responsible for kinetic mixing. The result differs significantly from the corresponding low-energy amplitudes for Higgs decays into $Z$ bosons. We use the low-energy theorem to obtain the branching ratios for exotic Higgs decays relevant for all models with charged baryon number. \\

\begin{acknowledgments}
\emph{\textbf{Acknowledgements.}}\,
The authors are grateful to \mbox{Joerg Jaeckel} for many helpful discussions on the topic. 
The authors would like to express special thanks to the Mainz Institute for Theoretical Physics (MITP) of the Cluster of Excellence PRISMA+ (Project ID 39083149), for its hospitality and support.
The work of MB and PF is supported by the UKRI Future Leaders Fellowship  DARKMAP.
\end{acknowledgments}

\bibliography{references}

\end{document}